\documentclass{article}
\usepackage{spconf,amsmath,graphicx, float}

\title{CiwaGAN: articulatory information exchange}
\name{Ga\v{s}per Begu\v{s}$^{1}$\thanks{ \textit{Corresponding author: Ga\v{s}per Begu\v{s} (begus@berkeley.edu).}}, Thomas Lu$^{1}$, Alan Zhou$^{2}$, Peter Wu$^{1\dagger}$, Gopala K. Anumanchipalli$^{1}$\sthanks{G.K.A.~and P.W.~are supported by NSF \#2106928.}}
\address{$^{1}$University of California, Berkeley, $^{2}$Johns Hopkins University%\\\texttt{\{begus,azhou314\}@berkeley.edu}
}
\usepackage{amsfonts}
\usepackage{amssymb}
\usepackage[safe]{tipa}
\usepackage[hidelinks]{hyperref}
\newcommand\rurl[1]{%
  \href{http://#1}{\nolinkurl{#1}}%
}
\begin{document}
\ninept
\maketitle
\begin{abstract}

Humans encode information into sounds by controlling articulators and decode information from sounds using the auditory apparatus. This paper introduces CiwaGAN, a model of human spoken language acquisition that combines unsupervised articulatory modeling with an unsupervised model of information exchange through the auditory modality. While prior research includes unsupervised articulatory modeling and information exchange separately, our model is the first to combine the two components. The paper also proposes an improved articulatory model with more interpretable internal representations. The proposed CiwaGAN model is the most realistic approximation of human spoken language acquisition using deep learning. As such, it is useful for cognitively plausible simulations of the human speech act.

\end{abstract}
\section{Introduction}
\label{intro}

Humans use sounds to exchange information. Sounds of speech are produced by moving articulators. During language acquisition, humans learn to control articulators  such that the resulting sound approximates the input language that they hear. At the same time, they need to learn to encode information into sounds and decode it from them. The learning process is thus highly complex and fully unsupervised: with the exception of visual stimuli from lips and tongue tip, language-acquiring children do not have a direct access to muscle activity or articulatory movements.

These two aspects of human speech process---information exchange and articulatory learning---have been modeled separately thus far. It has been shown that unsupervised deep learning models can learn to generate Electromagnetic articulography (EMA) channels in a Generative Adversarial Network (GAN) setting \cite{begusAGan}. On the other hand, a body of work shows that GAN-based models can learn to identify meaningful properties of human language in a fully unsupervised manner and use them to encode and decode information \cite{begus19,begusCiw,begus22Interspeech,begusLocal,begus2020identity}.

This paper combines the two proposals in Begu\v{s} \cite{begusCiw} and Begu\v{s} et al.~\cite{begusAGan} and tests whether a model that includes both information exchange and articulatory learning can learn linguistically meaningful properties in an interpretable way. The proposed model is the closest approximation of human language learning using deep learning approaches known to the authors. 

\section{The model}
\label{model}

\begin{figure*}
    \centering
    \includegraphics[width=.8\textwidth]{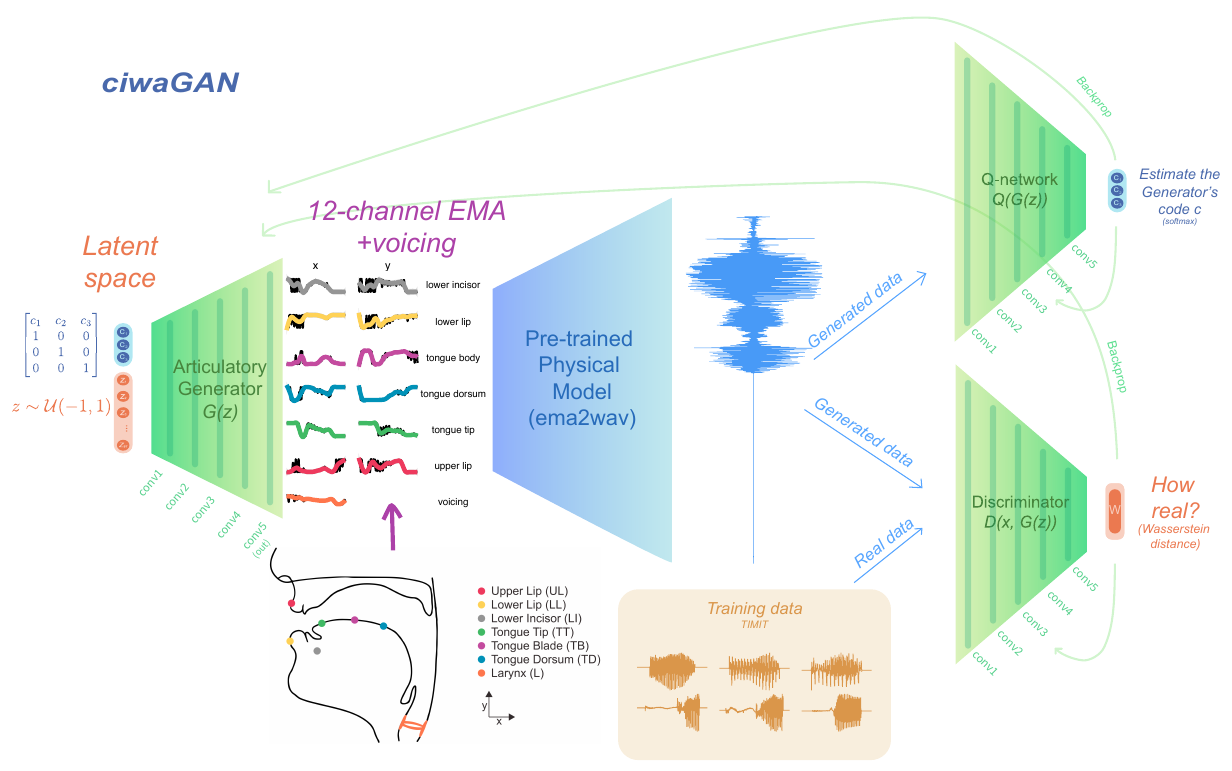}
    \caption{The architecture of the CiwaGAN (based on \cite{begusCiw,begusAGan,donahue19}).}
    \label{fig:icassp}
\end{figure*}

We propose a new unsupervised model of spoken language that combines the Articulation GAN architecture \cite{begusAGan} with the ciwGAN architecture \cite{begusCiw} into the \textit{CiwaGAN} proposal which stands for Categorical InfoWave Articulation GAN. Unlike in the Articulatory GAN proposal \cite{begusAGan}, the Articulatory Generator in ciwaGAN takes latent code variables $c$ in addition to uniformly distributed latent variables $z$ using which it generates 12 EMA channels and a channel for voicing. The $c$ variable is a one-hot vector.  A pre-trained physical model of sound production (ema2wav; \cite{wu22}) then turns the 12 EMA channels and the channel for voicing into waveforms.

The generated sounds are sent to the Discriminator and the Q-network. The Discriminator forces the Articulatory Generator to learn articulatory representations such that its outputs mimic real speech data. The Q-network, on the other hand, takes the generated audio and needs to decode the unique code from the Generator's latent space. The Discriminator thus mimics imitation and the Q-network mimics information exchange in human spoken language communication.

The structure of the Articulatory GAN and ema2wav is taken from \cite{begusAGan}, but  we use an improved unpublished ema2wav model \cite{wu22} with increased audio quality. We also introduce new hyperparameter choices that improve training: decreased stride and filter size which reduce noise and jitter in the EMA channels (Figure \ref{fig:12Ema}).

In addition to the original ArticulationGAN \cite{begusAGan} objective:

$$
\max_D \min_{G} V(D, G) = \mathbb{E}_{x \sim P_x}[D(x)] - \mathbb{E}_{z \sim P_z, c \sim P_c}[D(\mathcal{A}(G(z, c)))]
$$

we introduce an additional ``Q-network" following \cite{chen16} and \cite{begusCiw} which takes as input the waveform output of the ema2wav model and attempts to recover the latent code $c$ that was passed into the Articulatory Generator. Along with the Articulatory Generator, this network is optimized against the following ``Q-loss"  in an attempt to approximate the posterior distribution $Q(c|x)$:

$$\max_{G, Q} V_Q (G, Q) = \mathbb{E}_{x \sim \mathcal{A}(G(z, c))} \left[\mathbb{E}_{c' \sim P(c|x)} Q(c' | x)\right]$$

Putting this together, we get a new zero-sum game in which the Articulatory Generator is not only optimized to generate realistic gestures, but to do so in a way that its latent code $c$ is recoverable from the final auditory output:

$$\min_{G, Q} \max_D V(D, G) - V_Q(G, Q)$$

\begin{table}

 \scalebox{.69}{  \begin{tabular}{r|l|r|l}
        \hline\hline
        \multicolumn{2}{c|}{Articulatory Generator} & \multicolumn{2}{c}{Discriminator}\\
        \hline
        Layer &  Dimension & Layer & Dimension\\
        \hline
        $c, z$ & 100 $\times$ 1 & input & 20480 $\times$ 1\\
        fc + reshape & 16 $\times$ 1024 & conv0 & 5120 $\times$ 64\\ 
        upconv0 & 32 $\times$ 512 & conv1 & 1280 $\times$ 128\\ 
        upconv1 & 64 $\times$ 512 & conv2 &  320 $\times$ 256\\ 
        upconv2 & 128 $\times$ 256 & conv3 &  80 $\times$ 512\\
        upconv3 & 128 $\times$ 256 & conv4 & 20 $\times$ 1024\\
        upconv4 & 256 $\times$ 13 & flatten + logit & 1 $\times$ 1\\
        \hline\hline
    \end{tabular}}
    \centering
%   \vskip1em
   \scalebox{.69}{  \begin{tabular}{r|l}
        \hline\hline
        \multicolumn{2}{c}{Q-Network}\\
        \hline
        Layer &  Dimension\\
        \hline
        input & 20480 $\times$ 1\\
        conv0 & 5120 $\times$ 64\\
        qconv1 & 1280 $\times$ 128\\
        qconv2 &  320 $\times$ 256\\
        qconv3 &  80 $\times$ 512\\
        qconv4 & 20 $\times$ 1024\\
        flatten + logit & 9 $\times$ 1 \\
        \hline\hline
    \end{tabular}}

\caption{The structure of the Articulatory Generator (based on \cite{donahue19,begusCiw,begusAGan}). }
\end{table}

Prior work in articulatory speech synthesis primarily focuses on supervised methods with an objective to generate spoken language from articulatory data \cite{bocquelet14,aryal16,chen21,georges22,georges2020,wu22}. Siriwardena et al.~and Shamma et al.~\cite{mirrornet,mirrornet1,siriwardena23} propose an autoencoder architecture for unsupervised articulatory learning that does not model information exchange. The advantages of the GAN architecture \cite{goodfellow14,radford15,arjovsky17,donahue19} over autoencoders are that the Generator is trained by imitation and imagination rather than reproducing input data. From a cognitive modeling perspective, GANs are a more realistic approximation of human speech learning and our model  captures information exchange and articulatory learning simultaneously, which is not the case in previous models.

While GANs bring several advantages in modeling human spoken language learning, the new proposal also comes with several implementational challenges. The training objective is among the most challenging in unsupervised speech processing paradigm. Learning in the Articulatory Generator is fully unsupervised: the Generator needs to learn the 12 articulatory EMA channels + voicing based on the feedback from the Discriminator and the Q-network which get only audio inputs and no EMA data. In other words, the Generator network needs to learn to generate a completely different modality from what the Discriminator and Q-network receive for evaluations. At the same time, the Generator needs to learn to encode information into the generated data. The model never gets any kind of explicit information that would force it to learn lexical items. In principle, the Generator could encode any property of speech into the latent codes. Yet, given that pairing unique codes with lexical items is highly informative, the Generator predominantly learns to use these lexical items to convey information. Additionally, the ema2wav model is trained on a single speaker of British English, while the training data contains approximately 600 speakers of Standardized American English. Our model offers an advantage from the cognitive modeling perspective: humans learn to utilize their individual articulators based on auditory feedback from various individuals. However, this setting significantly increases the complexity of the training objectives. Finally, the learning is fully unsupervised in that the Generator and the Q-network never directly access the actual training data.

The improved ema2wav model (the ``Physical model'' in Figure \ref{fig:icassp}) is trained on the MNGU0 database \cite{richmond11} which consists of EMA recordings and corresponding waveforms of a single speaker of British English. This model is an updated version of Wu et al.~\cite{wu22} that initializes weights with those of a neural spectrum-to-waveform vocoder \cite{wu23}. This form of transfer learning improves the fidelity of the trained ema2wav model. The Discriminator is trained on nine TIMIT words: \textit{water, like, carry, greasy, ask, year, suit, dark, wash}, and between 600 and 700 distinct tokens are used for each word. These words were chosen because they were well-attested content words in TIMIT. The Generator is trained with a one-hot vector of length 9, corresponding to the 9 lexical items. It has been shown elsewhere that lexical learning happens even if the number of classes and the number of words do not match \cite{begusCiw}.  We train the model for 309,000 steps. For each Generator update, the Discriminator and the Q-network are updated five times. The entire code for the implementation of CiwaGAN is available at \url{https://github.com/gbegus/articulationGAN}.

\section{Results}

\subsection{Evaluation}

The performance of CiwaGAN is sufficiently high so that the outputs can be automatically evaluated (as opposed to being evaluated by a trained phonetician \cite{begusAGan}). We evaluate the outputs with a fine-tuned Whisper ASR model \cite{whisper}. 

The pretrained model Whisper-small \cite{whisper} is fine-tuned using a combination of TIMIT lexical items and CiwaGAN outputs manually annotated by the authors. The fine-tuning dataset consists of approximately 100 tokens from the TIMIT dataset for each of the 9 words used in training, combined with 800 samples annotated by the authors for a total of approximately 1,700 tokens. The model is fine-tuned for 250 steps and achieves a word error rate of 22.5\%. The entirety of the fine tuning dataset and all evaluations made by Whisper as well as the model's checkpoints are available at \url{http://doi.org/10.17605/OSF.IO/JBWYH}.

\begin{figure}
    \centering
    \includegraphics[width=.3\textwidth]{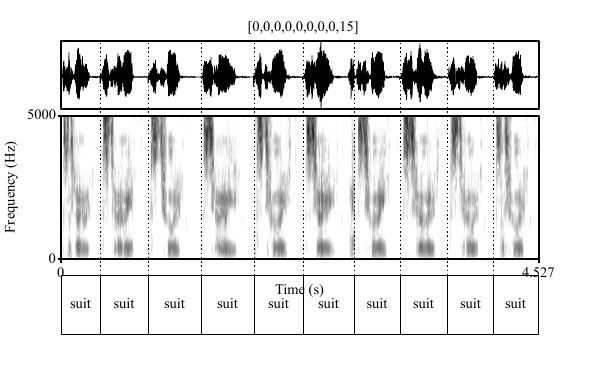}\hskip-1em
    \includegraphics[width=.3\textwidth]{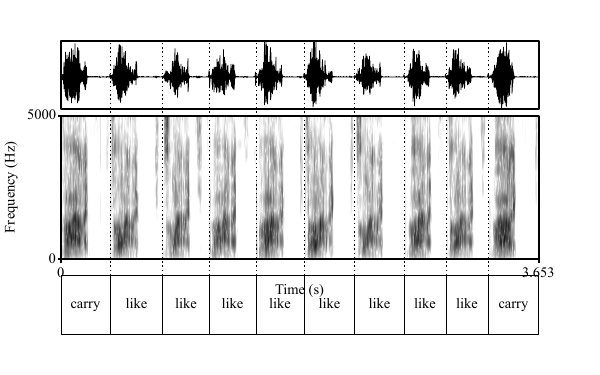}\\
    \includegraphics[width=.3\textwidth]{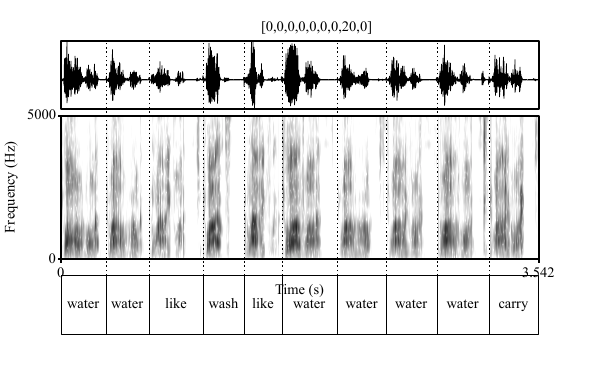}\hskip-1em
    \includegraphics[width=.3\textwidth]{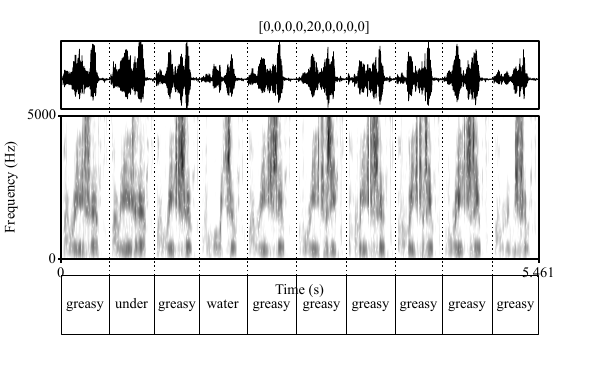}
    \caption{Ten cropped outputs independently sampled for four latent codes $c$ with individual variables set to extreme values (20 or 15). Under each waveform and spectrogram (0--5000 Hz) is a transcription by fine-tuned Whisper.}
    \label{fig:praat}
\end{figure}

\subsection{Evidence of information exchange}

\begin{figure}
    \centering
    \includegraphics[width=0.3\textwidth]{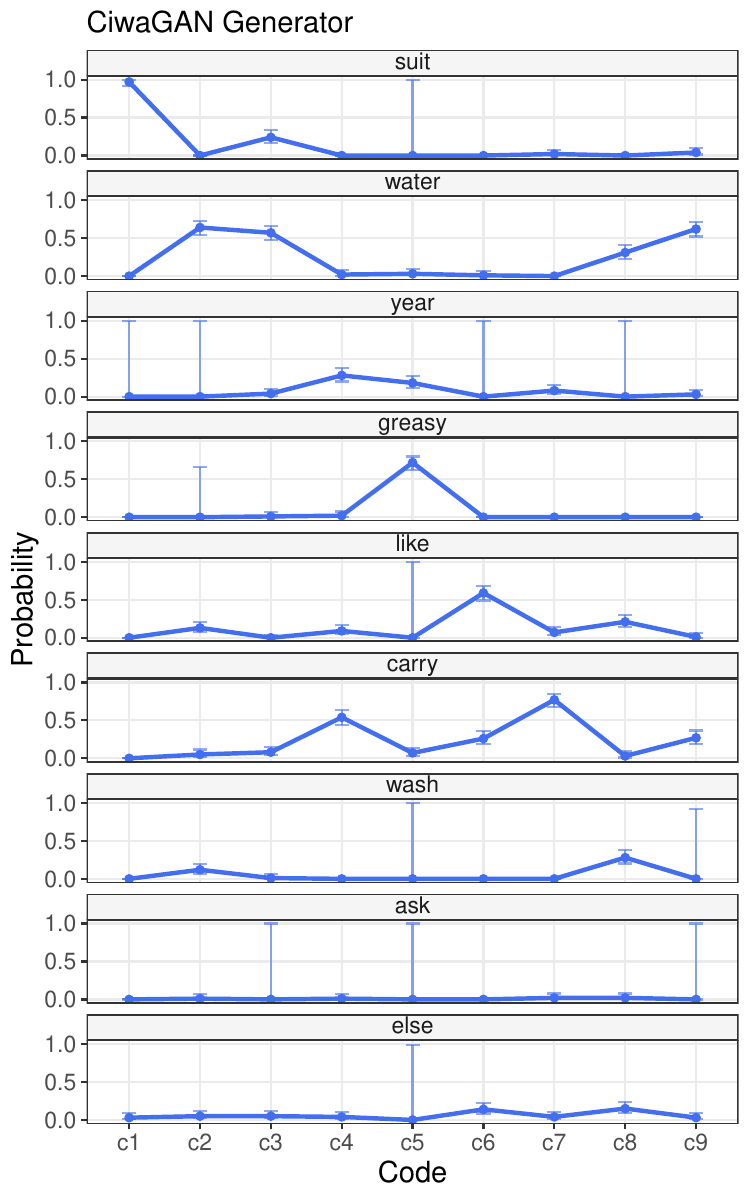}
    \caption{Estimates of the multinomial logistic regression model with Whisper transcriptions of the 900 generated words (100 per each category; grouped into the 9 words and an ``else'' condition for all other words) as the dependent variable and the code as the predictor when the value of each code is set to 20.}
    \label{fig:multinomial}
\end{figure}

To test whether the model learns to encode  linguistically meaningful information using articulatory representations, we utilize the technique for analyzing  underlying values of individual latent variables \cite{begus19,begusleban} by setting individual variables to extreme values outside the training range distribution. It has been shown that setting the latent codes to extreme values (e.g.~to [20,0,0]) when the training only contained latent codes with values of 0 or 1 (e.g.~[1,0,0]) results in a near-categorical output of a single lexical item in the CiwGAN architecture \cite{begusCiw}. In other words, this technique reveals the underlying learned representation for each latent code when individual latent variables are set to extreme values.

By applying the extreme value technique to the Articulatory Generator, we are testing the technique on a new frontier. By setting individual latent variables to extreme values, we can uncover the underlying representation of each unit and get a near categorical performance on lexical learning. Previous tests of this method have been limited to unimodal data. In this study, we are exploring underlying values of individual variables and their near-categorical performance when the generated data differs in modality from the input received by the Q-network: articulatory data vs.~auditory input.

Extreme values in CiwaGAN thus do not reveal the underlying audio words, but the underlying articulatory gestures that result in words which the Q-network learns to classify.

To quantify lexical learning, we generate 100 outputs for each one-hot code (but with values set at 20 instead of 1) while keeping the latent space constant across all 9 one-hot levels. This results in 900 outputs transcribed by fine-tuned Whisper. This data is then fit to a multinomial logistic regression using the \textit{nnet} package in R \cite{nnet} with the nine transcribed words (plus a category for other words ``else'') as the dependent variable and the unique code as the independent variable. The model with code as a predictor fits the data significantly better ($\text{df}=72, \text{AIC} = 1944.6$)
 than the model without this predictor ($\text{df}=8, \text{AIC} = 3518.8$).

Figure \ref{fig:multinomial} illustrates the extent of lexical learning in the CiwaGAN architecture. Most words display a single pronounced peak corresponding to a specific code when its value is set to 20 instead of 1. For example, with code set to [0,0,0,0,0,0,0,0,20], the network produces \textit{suit} 94 times out of 100 samples. Similarly, \textit{greasy} is produced 72 times (out of 100) with the code [0,0,0,0,20,0,0,0,0], and \textit{like} is transcribed 59 times for the code [0,0,0,0,0,20,0,0,0]. Other prominently encoded words include \textit{water}, \textit{year}, \textit{wash}, and \textit{carry}.  In contrast, words like \textit{ask} and \textit{dark} show no distinct peaks.

Figure \ref{fig:praat} demonstrates that the audio quality of words, generated using the extreme value technique, remains high even when individual code values significantly exceed the training range.

\subsection{Evidence of articulatory learning}

\begin{figure}
    \centering
    \includegraphics[width=.44\textwidth]{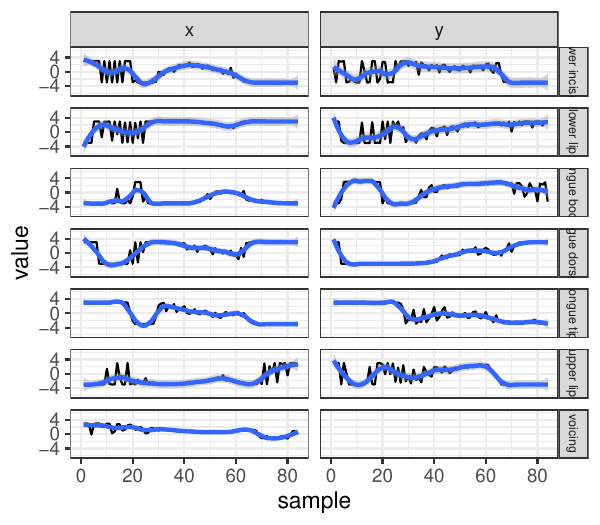}
    \caption{The 12 EMA channels with a channel for voicing generated by the Articulatory Generator for the sixth output (in Figure \ref{fig:praat}) of the code [0,0,0,0,0,0,0,0,15] transcribed as \textit{suit}. The blue line represents LOESS smoothing with a span of 0.2 used for calculating correlations in Table \ref{dtw}.} 
    \label{fig:12Ema}
\end{figure}

In the CiwaGAN, the Articulatory Generator learns specific articulatory gestures that result in individual words. By setting unique code values to extremes, we can induce the output of specific words. Consequently, CiwaGAN provides a framework for analyzing both lexical and articulatory learning.

To quantitatively assess articulatory learning, we perform a direct comparison between generated EMA recordings when the code is set to [0,0,0,0,0,0,0,0,15] which forces the output \textit{suit} (in 10/10 cases) and the actual EMA recording of the word \textit{suit} from the MNGU0 corpus. The generated EMA is LOESS-smoothed with a span of 0.2, yielding  highly interpretable articulatory gestures (Figure \ref{fig:12Ema}). With new hyperparameters that better model EMA data (lower filter size and stride), the EMA channels exhibit relatively minimal noise and jitter. Substantial jitter is evident during phases where specific articulators are not crucial to the articulation of the word. 

To quantify similarity between generated and real EMA, we perform Dynamic Time Warping (DTW) and conduct correlation tests on the DTW-aligned time series.

\begin{figure}
    \centering
    \includegraphics[width=.45\textwidth]{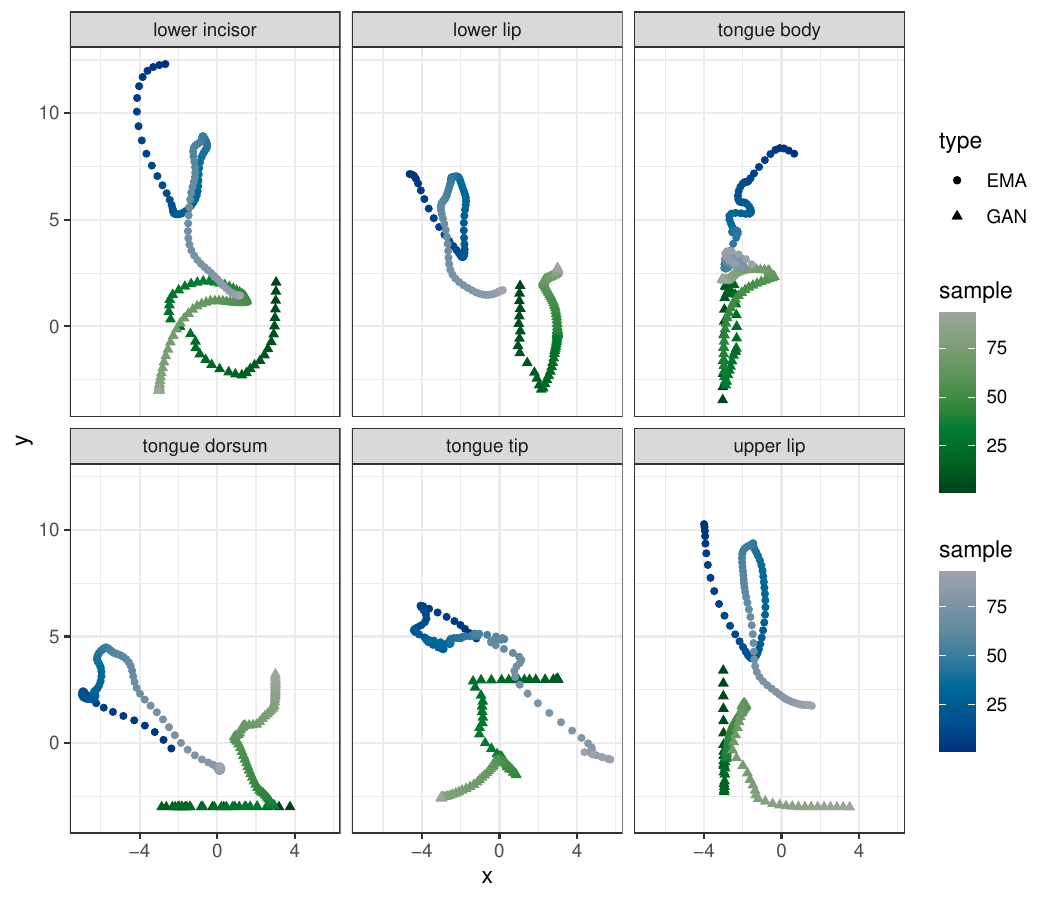}
    \caption{A comparison between generated (green triangle) and real EMA channels (blue circles) for a word \textit{suit}. The generated samples are from the sixth example of \textit{suit} in Figure \ref{fig:praat} when the code is set to [0,0,0,0,0,0,0,0,15] which generates \textit{suit} in 10 out of 10 trials. The real EMA data is taken from the MNGU0 database \cite{richmond11} and multiplied by a factor of 5 to match the magnitude of the generated EMA.}
    \label{fig:EMAGAN}
\end{figure}

The comparison between real and generated EMA (Table \ref{dtw} and Figure \ref{fig:EMAGAN}) shows that correlations are highest for articulators that are more relevant to the articulation of \textit{suit} such as the lips and tongue tip. In contrast, the tongue dorsum has a weaker correlation between the real and generated EMA. For the articulation of \textit{suit}, the positions of the lip (especially on the x-axis) and tongue tip position are crucial. The correlations for the tongue tip (on the y-axis) is  $r=0.94$. For the lower lip on the y-axis, it is $r=0.79$, while the upper lip has a correlation of $r=0.93$  on the x-axis and $r=0.86$ on the y-axis. 

Qualitatively, Figure \ref{fig:EMAGAN} illustrates that the articulatory gesture of the upper lip in the generated data closely mirrors that of the real EMA data, exhibiting a nearly identical looped pathway.

% latex table generated in R 4.3.0 by xtable 1.8-4 package
% Thu Sep  7 23:39:31 2023
\begin{table}
\centering
\scalebox{.8}{\begin{tabular}{lrrrr}
  \hline\hline
 Place & x.DTW  & x.Cor test & y.DTW  & y.Cor test \\ 
  \hline
 lower lip & 367.2 & 0.79 & 154.5 & 0.18 \\ 
 tongue tip & 168.4 & -0.43 & 149.3 & 0.94 \\ 
 lower incisor & 170.1 & -0.19 & 119.7 & 0.79 \\ 
 upper lip & 166.8 & 0.93 & 133.1 & 0.86 \\ 
 tongue body & 212.4 & -0.65 & 153.1 & -0.02 \\ 
 tongue dorsum & 182.7 & 0.84 & 217.9 & 0.04 \\ 
   \hline\hline
\end{tabular}}
\caption{The minimum global  DTW distance between smoothed generated EMA and real EMA using the \textit{dtw} package in R \cite{dtw}  and corresponding Pearson's product-moment correlations between the aligned time series for each of the 12 channels.
}
\label{dtw}
\end{table}

\section{Conclusion}

This paper proposes a new model that features several properties of human language: unsupervised learning, information exchange, articulatory learning, and the production-perception loop. We show that information exchange can occur even when the production model (the Articulatory Generator) needs to learn to generate a modality distinct from the perception network's modality: articulatory movements vs.~audio inputs. The model allows simulations of human speech activity that incorporate a realistic approach to information exchange alongside a detailed analysis of articulatory gesture learning. The proposed  model thus represents one of the closest modeling approximations of human language acquisition. 

\clearpage

\bibliographystyle{IEEEbib}
\bibliography{bibliography.bib}

\end{document}